\documentclass[12pt]{article}%
\usepackage{graphics}%
\begin{document}%
%--------------------------------------------------------------
%--------------------------------------------------------------
%\def\d{{\mathrm{d}}}
\def\e{{\mathrm e}}%
\def\g{{\mbox{\sl g}}}%
\def\Box{\nabla^2}%
\def\d{{\mathrm d}}%
\def\R{{\rm I\!R}}%
%--------------------------------------------------
\def\ie{{\em i.e.\/}}%
\def\eg{{\em e.g.\/}}%
\def\etc{{\em etc.\/}}%
\def\etal{{\em et al.\/}}%

%--------------------------------------------------------------
\markright{Cosmology as a search for overall equilibrium\hfil}%
%--------------------------------------------------------------
\title{\bf \LARGE Cosmology as a search  \\ for overall equilibrium}%
%--------------------------------------------------------------
\author{Carlos Barcel\'o%
\\[2mm]%
%-----------------------------------
{\small\it%
\thanks{\tt carlos@iaa.es}%
\ Instituto de Astrof\'{\i}sica de Andaluc\'{\i}a, CSIC,
Camino Bajo de Hu\'etor 50, 18008 Granada, Spain}%
\\[4mm]%
}
%--------------------------------------------------------------
\date{{\small 29 October 2006; {\LaTeX-ed \today}; gr-qc/0611090}}%
%--------------------------------------------------------------
\maketitle%
%--------------------------------------------------------------
\begin{abstract}%

In this letter we will revise the steps followed by A. Einstein when
he first wrote on cosmology from the point of view of the general
theory of relativity. We will argue that his
insightful line of thought leading to the introduction of the cosmological
constant in the equations of motion has only one weakness: The constancy
of the cosmological term, or what is the same, its independence
of the matter content of the universe. Eliminating this feature,
I will propose what I see as a simple and reasonable modification of the 
cosmological equations of motion. The solutions of the new cosmological 
equations give place to a cosmological model that tries to approach
the Einstein static solution. This model shows very appealing
features in terms of fitting current observations.

%--------------------------------------------------------------
\vspace*{5mm}%
\end{abstract}%

%\clearpage

%--------------------------------------------------------------
\section{Introduction}%
\label{sec:intro}%
\setcounter{equation}{0}%
%--------------------------------------------------------------

The way Einstein arrived at his formulation of a cosmological model
for the universe was based on two basic
hypothesis~\cite{einstein-universe}. From the one side, he was
struggling with the idea that a single particle in the Universe
couldn't have inertia as inertia should have a relational origin. From
the other side, he thought that the universe could be described as a
gas of stars close to equilibrium. After some intricate reflections,
he arrived to the notion of a finite universe without boundary, and 
static on the overall, that is, at equilibrium. For that he introduced
the now famous cosmological constant as a way of compensating the
purely-attractive behaviour of matter in his equations.

In 1930 A. Eddington proved that Einstein's static Universe was
unstable under homogeneous departures from the equilibrium
state~\cite{eddington}. This together with the observation of the
recession of galaxies led progressively to the abandonment of Einstein
model. Anyway, the Einstein model has been the subject of several
investigations over the years [see for example 
\cite{lemaitre}-\cite{gibbons}]. The most recent is a revision 
of its possible role as the initial state for a past eternal 
universe~\cite{barrow}.

Our point of view in this letter is that the reasoning that led
Einstein to his model is too powerful to be simply abandoned. Of
course, from nowadays perspective, his assumption of a universe at
precise overall equilibrium clearly needs some revision. But more
importantly, one needs to revise his way of incorporating an
equilibrating mechanism by adding a cosmological constant.

At that time, physicist didn't seem to suspect that the cosmological
term could be just another type of manifestation of matter, what we
now use to call a vacuum-energy contribution\footnote{At least it
seems that it was not a popular idea. Remarkably, the first account of
the possible existence of zero-point energy contributions can be
traced back to Nerst in 1916~\cite{nerst}. However, one could say that
the idea of a connection between zero-point energy and cosmological
constant was not popularized up to Zeldovich's
work~\cite{zeldovich}.}. For example, now we know that the presence of
any quantum field in nature could give place naturally to a
vacuum-energy contribution with an equation of state of the form
$p_V=-\rho_V$. However, when the quantum fields are considered as
fundamental, the naive calculation of this vacuum-energy contribution
is always huge, being completely at odds with the observations (apart
from being completely independent of the total amount of energy in the
non-vacuum component of the field).

In recent years we have learnt, mostly through the works of
G. Volovik~\cite{volovik-book,volovik-paper}, that the universe that
we experience might be an emergent structure ``living'' at the
low-energy corner of a fundamental quantum system with a finite number
of degrees of freedom, similar somehow to a quantum
liquid. Remarkably, in these systems the set of excitations with
respect to the vacuum state include collective-excitations,
describable in terms of geometrical (gravitational) an electromagnetic
fields, and quasi-particle excitations, describable as quantum matter
(fermionic) fields. A very interesting insight coming out from the
observation of these systems is that, the calculation of a vacuum
energy contribution performed by adding the zero-point
energies associated to the quantization of the different
quasi-particles modes leads to grossly incorrect
results. Vacuum-energy has to be calculated by other means.

It is not difficult to understand that for a condensed matter system
to have an equilibrium vacuum state in the absence of external forces
its vacuum pressure has to vanish, $p_V=0$ . This is exactly what
happen with quantum liquids. Now, if these type of systems are brought
out of their lowest energy state by exciting, for example, a thermal
cloud of quasi-particles within them, they tend to reach a new state
of internal equilibrium. The pressure associated with the
quasi-particles, $p_M$, is equilibrated by a newly generated vacuum
term $p_V$ such that $p_M+p_V=0$. In this way we see that the vacuum
term is not constant, but depends on the characteristics of the matter
contribution, tending to counterbalancing it. The exact counterbalance
occurs only at precise equilibrium. If these systems were set up out
of equilibrium, they would evolve towards the equilibrium
point. During this transient regime one would have $p_M+p_V \neq 0$.

Another line of thought, this time exclusively within the context of
gravitational interactions, that also suggest the existence of an
equilibrating mechanism is the following. Einstein gravitational law,
expressed as $G_{\mu\nu}=8\pi G T_{\mu\nu}^M$, has been seen to work
to a high degree of accuracy in the astronomical (as opposed to
cosmological) context\footnote{By Einstein gravitational law I mean
not only that Einstein tensor is equal to an energy-momentum tensor,
but that there is a way of prescribing the form of the energy-momentum
tensor given some local (Minkowskian) notion of mass.} (in the weak
field limit they precisely give place to Poisson's equation $\nabla^2
\phi=\rho_M$ for the Newtonian potential $\phi$). However, if the
universe was on the overall a three sphere (finite without boundary),
one would expect in Newtonian terms that a quantity of localized
matter will affect another quantity of localized matter with two
direct contributions, associated with the shortest and the longest
geodesic path between the two masses, plus additional contributions
associated with geodesic paths connecting them after having traverse
entirely the universe one or several times. Some of these
contributions are attractive (as the one associated with the shortest
direct path between the masses) but some other are effectively
repulsive (as the one associated with the longest direct path). As a
result, two masses at a fixed distance will affect each other
differently depending on the size of the universe. For example, in a
extreme situation, two masses located at antipodal points of a closed
universe would exert no net-force into each other. We can symbolically
write the global result of all these gravitational interactions as a
modified Poisson equation
\begin{eqnarray}
\nabla^2 \phi=\rho_I(\phi,\rho_M)\;,
\end{eqnarray}
where $\rho_I$ identifies an effective matter source depending on the
real matter density but also on the geometry of the universe itself
(here represented only by the scalar potential)\footnote{Einstein
contemplated a simple modification of the Poisson equation in the form
of the so-called Seeliger equation $\nabla^2\phi=-\Lambda
\phi + \rho_M$.}. Thinking in this way, it is not difficult to reach
the conclusion (as Einstein did) that Einstein equations should be
modified to deal with cosmological situations. These modified
equations should contain cosmological solutions (homogeneous and
isotropic) that are static. The easiest modification one can think of
that do the job is one in which the global effects of matter are
separated into a normal contribution and a cosmological contribution
equilibrating each other -- this is precisely Einstein
universe. However, there is no reason to adopt the much stronger
assumption that the cosmological term is constant.

The possible existence of a varying cosmological term was considered
as early as in 1933 by M. Bronstein~\cite{bronstein} (see for
example~\cite{peebles} and~\cite{padmanabhan} for reviews of the
history of the cosmological term). Nowadays there are many models
incorporating a varying cosmological term (mainly under the names of
quintessence~\cite{caldwell} and dark energy~\cite{turner}; see
also~\cite{peebles,padmanabhan} and references therein for a summary
of different models). Generically, the idea has been that since
inflation (or some other phase transitions in the early universe) the
contribution of an effective cosmological term has been progressively
disappearing, leaving a small or zero contribution at present. The
cosmological contribution has been usually modelled as an independent
field.  Here we are going to treat the cosmological contribution as
completely tied up with the normal matter content.

In the following we will propose a simple heuristic modification of
the cosmological equations of motion incorporating an equilibrating,
and so variable, cosmological term. Adjustment mechanisms have been
explored before (see~\cite{dolgov} for an updated discussion). The
difference here is that instead of describing a specific and detailed
adjusting mechanism an see what happens, we take as an starting
hypothesis the very existence of an equilibrating mechanism. The
equations that we propose to implement this mechanism are arguable the
simplest one can think of, and are enough to illustrate the main
features of these type of models.

%--------------------------------------------------------------
\section{The model}%
\label{sec:intro}%
\setcounter{equation}{0}%
%--------------------------------------------------------------

Let us assume that it exists a state of stable internal (by definition
there is nothing outside the Universe) equilibrium for the universe on
the overall . To describe the situation let us separate the total
effect of matter on the geometry as a normal contribution and a
so-called vacuum contribution. For concreteness and simplicity let us
choose an equation of state for matter of the form $p_M= (\gamma
-1)\rho_M$ with $\gamma \in (4/3, 1)$. The equation of state for the
vacuum energy is $p_V=-\rho_V$.

The Friedman equation for a universe with closed spherical 
spacial sections is  
\begin{eqnarray}
\left({\dot a \over a}\right)^2 = -{1\over a^2}+{8\pi G \over 3}\rho_T\;,
\label{first}
\end{eqnarray}
where $\rho_T$ contains the matter density and the vacuum energy
contribution: $\rho_T= \rho_M+\rho_V$.  The equation for the 
acceleration of the scale factor is 
\begin{eqnarray}
{\ddot a \over a}=-{4\pi G \over 3}(\rho_T +3p_T)\;.
\end{eqnarray}
The Einstein static universe is obtained when 
\begin{eqnarray}
(\rho_T +3p_T)=0\;, ~~~~~~ {8\pi G \over 3} \; a^2\rho_T =1\;
\end{eqnarray}
and $a=a_s$ equal constant. In this case 
\begin{eqnarray}
\rho_V= {3 \gamma -2 \over 2} \rho_M
~~~{\rm and}~~~ a_s^{-2}={8\pi G \over 3} {3 \gamma \over 2} \rho_M\;.
\end{eqnarray}

Now let us eliminate the condition $\rho_V:=$constant. 
This condition is obtained if one requires that the conservation 
of the energy-momentum tensor applies separately to the normal-matter
and vacuum-energy components. When this is not the case, from
the total conservation equation
\begin{eqnarray}
{d \rho_T \over da}= - {3 \over a}(\rho_T+p_T)\;,
\end{eqnarray}
one obtains
\begin{eqnarray}
{d \rho_V \over da}= -\left({d \rho_M \over da} + 
{3\gamma \over a}\rho_M \right)\;.
\label{second}
\end{eqnarray}
This equation and Friedman's equation have to be supplemented with a
third equation to obtain a closed set of equations. The third
equation will be the one that regulates the transference of energy between
the normal and the vacuum energy components. A simple choice one
can make is
\begin{eqnarray}
{d \over da}\left[\rho_V- {(3\gamma -2) \over 2} \rho_M \right]
= -{1\over \tau} \; \rho_V\;.
\label{third}
\end{eqnarray}
We will show that this nicely incorporates the idea that the universe
has the tendency to go towards the equilibrium point\footnote{ This
equation is in conceptual tune with what happen in a quantum field
theory over a curved background: Acceleration of the universe has the
tendency to creating particles from the vacuum, lowering through
back-reaction the vacuum energy density. Combining the equation of
conservation~(\ref{second}) and the equation for
interchange~(\ref{third}), one obtains ${d \rho_M \over da} = -{2
\over a} \rho_M + {2 \over 3 \gamma}{1 \over \tau}\rho_V$. The first
term represents an effective dilution of $\rho_M$ due to the expansion
and the second term could be interpreted as an increase of the matter
density owing to particle creation from the vacuum. However, the
similarity is just conceptual, no quantitative.}. The constant
parameter $\tau$ can be interpreted as a relaxation time and sets the
strength of the tendency towards the equilibrium: As smaller is $\tau$
the greater is this tendency. At this stage, we don't have a
fundamental theory from which to calculate the value of $\tau$, we
just take it as a parameter of the model.

Let us now first solve equations~(\ref{second}) and (\ref{third}). The
general solution will be expressed in terms of two arbitrary constants
$C_1$ and $C_2$. Once found the functions $\rho_V=\rho_V(a,C_1,C_2)$
and $\rho_M=\rho_M(a,C_1,C_2)$ we can plague them into the Friedman
equation to solve for the scale factor. Friedman's equation,
(\ref{first}), can be written as
\begin{eqnarray}
{1 \over 2} \dot a^2 + V_{\rm Newt-eff}(a) = 0\;,
\label{newtonian}
\end{eqnarray}
with
\begin{eqnarray}
V_{\rm New-eff}(a) \equiv {1 \over 2}
\left\{1 - {8\pi G \over 3}a^2[\rho_V(a)+\rho_M(a)] \right\}\;.
\end{eqnarray}
In this form it is very easy to extract its physical information.

So, from equations (\ref{second}) and (\ref{third}) we obtain 
a second order equation for $\rho_M$:
\begin{eqnarray}
{d^2 \rho_M \over da^2}
+\left({2 \over a}+ {2 \over 3\gamma}{1 \over \tau}\right){d \rho_M \over da}
+\left({2 \over a\tau}-{2 \over a^2}\right)\rho_M.
\label{second-order}
\end{eqnarray}
The general solution of this second order differential equation is 
\begin{eqnarray}
\rho_M(a)=&&\hspace{-6mm}C_1 a^{-1} 
\exp\left(-{1  \over 3\gamma} {a  \over \tau}\right) \;
M_{3\gamma-1,3/2}\left({2 \over 3\gamma} {a \over \tau} \right)
\nonumber
\\
&&\hspace{-6mm}+C_2 a^{-1} 
\exp\left(-{1  \over 3\gamma} {a  \over \tau}\right) \;
W_{3\gamma-1,3/2}\left({2 \over 3\gamma} {a \over \tau} \right)\;,
\end{eqnarray}
where $M_{\kappa,\lambda}$ and $W_{\kappa,\lambda}$ identify Whittaker's 
functions~\cite{abramowicz-stegun}.
Now one can also solve for $\rho_V(a)$: 
\begin{eqnarray}
\rho_V={3\gamma \over 2} \;\tau \;{d \rho_M \over da}
+3\gamma \; {\tau \over a} \; \rho_M \;.
\end{eqnarray}
A remarkable fact to notice is that although there are two
independent solutions to previous equation, only $M_{3\gamma-1,3/2}$ is 
different from zero for $a>0$. Therefore concerning the behaviour of the 
different magnitudes for positive $a$, only the value of $C_1$ is relevant.

\begin{figure}[htb]
\vbox{
\hfil
\scalebox{0.5}{\includegraphics{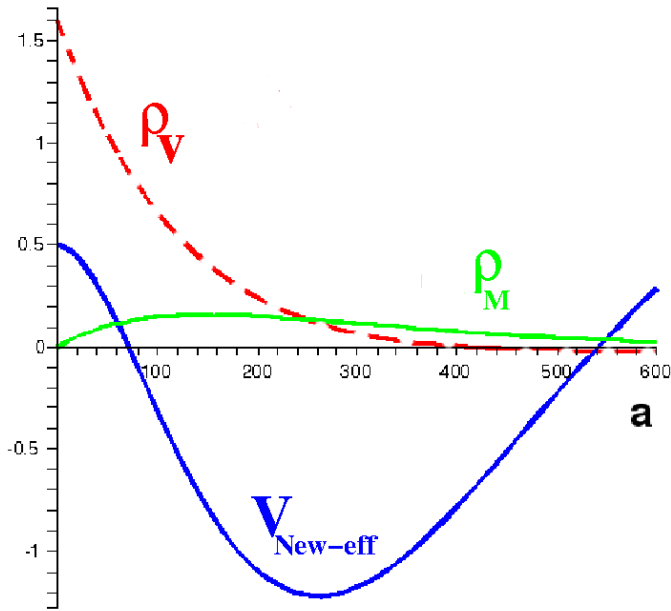}}
\hfil
}
\bigskip
\caption{\label{Fig:rhoV-rhoM-V} Plots of $\rho_V(a)$, $\rho_M(a)$ and
$V_{\rm Newt-eff}(a)$ showing a generic solution of our set of cosmological
equations. We have used $\gamma=1.25$ $C_1=100$, $C_2=0$,
$\tau=100$ and $8\pi G/3=0.0002$ for graphical clarity.}
\end{figure}

In figure~\ref{Fig:rhoV-rhoM-V} we have plotted a generic behaviour of 
$\rho_V(a)$, $\rho_M(a)$ and $V_{\rm New-eff}(a)$. We have chosen 
$C_1=100$, $C_2=0$, $\gamma=1.25$, $\tau=100$ and $8\pi G/3=0.0002$ 
for graphical clarity.
%====================================================
%====================================================
The red dashed line, the green bold line and the blue bold line
represent $\rho_V$, $\rho_M$ and $V_{\rm New-eff}$ respectively.  We
clearly observe that the previous set of equations give place
generically to oscillating universes between a minimum and a maximum
size [remember that the Newtonian effective energy is equal zero;
see~(\ref{newtonian})]. The Einstein solution is a fine-tuned solution
corresponding to the situation in which the bottom of the potential is
precisely at the zero-energy level. However, within this framework
Einstein's solution would be stable\footnote{G. Volovik and the author
already pointed out this possibility in a scenario based on emergent
gravity under the influence of an external thermal
bath~\cite{barcelo-volovik}.}. The generic cosmological solutions
found are such that at a large value of the vacuum energy density the
universe is small but not singular; at that point the matter energy
density is small. Then, as the universe expands, the vacuum energy
density rapidly decreases while the matter energy density
increases. In this process the matter energy density becomes
sufficiently large so that, at some maximum size, the system stops its
expansion being driven to re-collapse towards the original
point. Then, the cycle is repeated over and over. Oscillating models
caused by a cosmological term diminishing with the expansion have been
consider before in~\cite{overduin}. In that paper instead of
introducing as here an equation of transference of energy, they
analyse different functional forms for the dependence of $\rho_V$ with
the scale factor.

The model has only two adjustable parameters: The time relaxation parameter
$\tau$ and the initial value of $\rho_M$. The initial value of
$\rho_V$ is then also fixed. This reflects the fact that we are really
not dealing with two matter sources but with a single one effectively
described as separated into two terms.

%--------------------------------------------------------------
\section{Cosmological observations}%
\label{sec:intro}%
\setcounter{equation}{0}%
%--------------------------------------------------------------

The presented oscillating universe only considers the homogeneous and
isotropic degree of freedom of the gravitational and matter
sectors. In a more realistic setting, the energy on the homogeneous
and isotropic mode would decrease owing to the transfer of energy to
inhomogeneous modes. Then, one can define an entropic birth for the
universe as the starting point of a cycle at which all of the
inhomogeneous modes were unexcited. In terms of the cosmological time,
the universe has not a beginning; instead the universe is past
eternal. Therefore, if one takes any snapshot in the evolution having
some inhomogeneities present and, then, evolve backwards in time, one
will always find a cosmological time in the past at which all the
inhomogeneous modes were unexcited. This ``initial'' cycle would have
the largest amplitude (that is, the largest difference between $a_{\rm
min}$ and $a_{\rm max}$). Then, each new cycle would have a smaller
amplitude than the previous. Eventually, the homogeneous mode of the
universe would settle down at the Einstein equilibrium point.%
%In fact, we would expect this mechanism to be quite efficient
%so that the oscillations would be quickly damped. 

The validity of a cosmological model can only be assessed by the
observations. Thus, this model will have to surpass several tests
in the future. At this stage I can only say that it has very
appealing features including its falsifiability:
\begin{itemize}

\item
It is reasonable to think that we are already leaving in a world of
low-energies (all the phenomena around us happen at energy scales much
lower that Planck scale). It is also reasonable to think that we are
neither almost at the equilibrium point (in the cosmological sense)
nor close to the initial state.  In terms of entropy, there is already
bast amounts of entropy in the universe, but clearly, we are very far
from a maximal entropy state. Therefore, starting from this
observation our model tell us that the values of $\rho_M$ and $\rho_V$
should be of the same order. This so-called coincidence is something
that we observe and has been part of the motivation for the present
analysis.

\item
Taking the currently accepted values of $\Omega_\Lambda \sim 0.7$ and
$\Omega_M \sim 0.3$~\cite{cosmological-parameters}, our model predicts
that the acceleration that we observe should be decreasing with time
at present and so be bigger in the past. The fitting of the supernova
data at high redshifts seems to provide an indication of the
contrary~\cite{riess}. The acceleration appears as nonexistent in the
past, giving support to models of dark energy as the Chaplygin gas
(see for example~\cite{bertolami}). However, by making global
comparisons between different cosmological models, other authors argue
that it is still impossible to discern, for example, between a
cosmological constant and varying dark energy~\cite{liddle}.

Another prediction of this model is that the normal-matter
energy density at present and at the recent past would have to have a
different evolution that in standard cosmological models. For a dust
dominated universe ($\gamma=1$) one only has to compare the standard
and modified behaviours:
\begin{eqnarray}
{d \rho_M \over da}=-{3 \over a} \;\rho_M ~~~~{\rm versus}~~~~ 
{d \rho_M \over da} = -{2 \over a} \; \rho_M + 
{2 \over 3 \gamma}\;{1 \over \tau} \; \rho_V \;.
\end{eqnarray}

\item
The ``close to Big Bang'' origin of the universe in our model allows
for incorporating most of the predictive power of this
paradigm. However one has to bear in mind that any realistic
calculation within our model will have to take into account two new
factors: i) Since its entropic birth, the universe could have passed
through a few entire cycles before entering in its current expansive
phase; ii) In each new cycle, the maximum value of the acceleration
attained, proportional to $\rho_V-[(3\gamma-2)/2]\rho_M$, would be
smaller.

For example, the diminishing of the duration of a single phase of 
nucleosynthesis, owing to the background acceleration predicted for that 
period, could be compensated with the plausible existence of a
few cycles reaching large enough temperatures for nuclear reactions to
take place.

\item
In this model the time elapsed since the entropic birth of the
universe (Big Bang-like) would be much larger than in standard
cosmological models. We have much more time to produce structures in
the universe, something that has always been problematic in standard
cosmology. 

\item
This model does not have the so-called horizon problem of standard
cosmology as there is not a ``beginning of time'' event. Therefore,
radiation could have enough time to thermalize in very large
scales. If the current cycle would have started at a temperature
smaller than $T_{\rm recombination}$, then, the size of the
inhomogeneities found in the cosmic microwave radiation would not
directly constraint the size of the inhomogeneities in the barionic
matter sector in recent times.

\item
Within this positive curvature model, the found values $\Omega_\Lambda
+\Omega_M \sim 1$ and $H_0^{-1}\sim 14.000$ Gyears would be just
telling us that the universe is very large, $R_0 \geq 100$ Gpc with
$R_0$ its current physical radius, so locally it would be almost flat.

\end{itemize}

Further checking of the compatibility of this cosmological model with 
actual observations will be the subject of future work.

%------------------------------------------------------------------
\section*{Acknowledgements}
%----------------------------------------------------------------- 

I would like to thank Grisha Volovik for setting up with his questions
and suggestions some of the seeds of this work. I also would like to
thank Narciso Ben\'itez, Stefano Liberati, Jos\'e Luis Jaramillo and
Mariano Moles for very useful comments.

%------------------------------------------------------------------

%------------------------------------------------------------------------------

%---------------------------------------------------------------------

\begin{thebibliography}{99}
%----------------------------------------------------------------- 


%--------------------------------------------------------------
\bibitem{einstein-universe}%
A.~Einstein,
%``Kosmologische Betrachtungen zur allgemeinen Relativit\"atstheorie,''
Sitzungsberichte der K{\"o}niglich Preussischen Akademie der 
Wissenschaften 142, (1917);
Also in a translated version in {\it The principle of Relativity},
(Dover, New York, 1952).
%--------------------------------------------------------------
\bibitem{eddington}
A.~S.~Eddington,
%``On the instability of Einstein's spherical world,''
Mon.\ Not.\ Roy.\ Astron.\ Soc.\  {\bf 90}, 668 (1930).
%%CITATION = MNRAA,90,668;%%
%----------------------------------------------------------------
\bibitem{lemaitre}
G.~Lema\^itre,
%''Expansion of the universe,''
Mon.\ Not.\ Roy.\ Astron.\ Soc.\  {\bf 91}, 490 (1931).
%%CITATION = MNRAA,91,490;%%
%----------------------------------------------------------------
\bibitem{bonnor}
W.~B.~Bonnor,
%``The instability of the Einstein universe,''
Mon.\ Not.\ Roy.\ Astron.\ Soc.\  {\bf 115}, 310 (1954).
%%CITATION = MNRAA,115,310;%%
%----------------------------------------------------------------
\bibitem{harrison}
E.~R.~Harrison,
%''Normal Modes of Vibrations of the Universe,''
Rev.\ Mod.\ Phys.\  {\bf 39}, 862 (1967).
%%CITATION = RMPHA,39,862;%%
%----------------------------------------------------------------
%\cite{Gibbons:1987jt}
\bibitem{gibbons}
G.~W.~Gibbons,
%``The Entropy And Stability Of The Universe,''
Nucl.\ Phys.\ B {\bf 292}, 784 (1987).
%%CITATION = NUPHA,B292,784;%%
%----------------------------------------------------------------
%\cite{Barrow:2003ni}
\bibitem{barrow}
J.~D.~Barrow, G.~F.~R.~Ellis, R.~Maartens and C.~G.~Tsagas,
%``On the stability of the Einstein static universe,''
Class.\ Quant.\ Grav.\  {\bf 20}, L155 (2003)
[arXiv:gr-qc/0302094].
%%CITATION = GR-QC 0302094;%%
%----------------------------------------------------------------
\bibitem{nerst}%
W. Nerst, Verh. Dtsch. Phys. Ges. {\bf 18}, 83 (1916)
%----------------------------------------------------------------
\bibitem{zeldovich}%
Ya. B. Zel'dovich, 
%''Cosmological Constant and Elementary Particles,''
Zh. Eksp. Teor. Fiz., Pis'ma Red. 6, 883 (1967) [JETP Lett. 6, 316 (1967)].
%----------------------------------------------------------------
\bibitem{volovik-book}%
G.~E.~Volovik,
``The Universe in a Helium Droplet,''
(Clarendon Press, Oxford, UK, 2003).
%\href{http://www.slac.stanford.edu/spires/find/hep/www?irn=5651816}{SPIRES entry}
%--------------------------------------------------------------
\bibitem{volovik-paper}%
%\cite{Volovik:2006bh}
%\bibitem{Volovik:2006bh}
  G.~E.~Volovik,
  ``Vacuum energy: Myths and reality,''
  arXiv:gr-qc/0604062;
  %%CITATION = GR-QC 0604062;%%
\\
%\cite{Volovik:2004gi}
%\bibitem{Volovik:2004gi}
%  G.~E.~Volovik,
%  ``Cosmological constant and vacuum energy,''
  Annalen Phys.\  {\bf 14}, 165 (2005).
  [arXiv:gr-qc/0405012].
  %%CITATION = GR-QC 0405012;%%
%--------------------------------------------------------------
\bibitem{bronstein}
M. Bronstein, 
Phys. Z. Sowjetunion 3, 73 (1933).
%--------------------------------------------------------------
%\cite{Peebles:2002gy}
%\bibitem{Peebles:2002gy}
\bibitem{peebles}
  P.~J.~E.~Peebles and B.~Ratra,
  %``The cosmological constant and dark energy,''
  Rev.\ Mod.\ Phys.\  {\bf 75}, 559 (2003)
  [arXiv:astro-ph/0207347].
  %%CITATION = ASTRO-PH 0207347;%%
%--------------------------------------------------------------
%\cite{Padmanabhan:2002ji}
%\bibitem{Padmanabhan:2002ji}
\bibitem{padmanabhan}
  T.~Padmanabhan,
  %``Cosmological constant: The weight of the vacuum,''
  Phys.\ Rept.\  {\bf 380}, 235 (2003)
  [arXiv:hep-th/0212290].
  %%CITATION = HEP-TH 0212290;%%
%--------------------------------------------------------------
\bibitem{caldwell}
R.~R.~Caldwell, R.~Dav\'e and P.~J.~Steinhardt, 
%''Cosmological Imprint of an Energy Component with General Equation of State,'' 
Phys. Rev. Lett. {\bf 80}, 1582 (1998).
%--------------------------------------------------------------
\bibitem{turner}
M.~S.~Turner, in The Galactic Halo, Astronomical Society
of the Pacific Conference Proceedings No. 165, edited by
B. K. Gibson, T. S. Axelrod, and M. E. Putnam (Astronomical
Society of the Pacific, San Francisco 1999).
%--------------------------------------------------------------
%\cite{Dolgov:2005bi}
%\bibitem{Dolgov:2005bi}
\bibitem{dolgov}
  A.~D.~Dolgov,
%   ``Problems of cosmological constant, dark energy and possible adjustment mechanism,''
  Int.\ J.\ Mod.\ Phys.\ A {\bf 20}, 2403 (2005).
  %%CITATION = IMPAE,A20,2403;%%
%--------------------------------------------------------------
\bibitem{abramowicz-stegun}%
M. Abramowicz and I. A. Stegun, Handbook of Mathematical Functions 
(Dover, New York, 1972)
%----------------------------------------------------------------
\bibitem{barcelo-volovik}
%\cite{Barcelo:2004he}
%\bibitem{Barcelo:2004he}
C.~Barcel\'o and G.~Volovik,
% ``A stable Einstein universe,''
JETP Lett.\  {\bf 80}, 209 (2004)
[Pisma Zh.\ Eksp.\ Teor.\ Fiz.\  {\bf 80}, 239 (2004)]
[arXiv:gr-qc/0405105].
%%CITATION = GR-QC 0405105;%%
%----------------------------------------------------------------
%\cite{Overduin:1998zv}
%\bibitem{Overduin:1998zv}
\bibitem{overduin}
  J.~M.~Overduin and F.~I.~Cooperstock,
  %``Evolution of the Scale Factor with a Variable Cosmological Term,''
  Phys.\ Rev.\ D {\bf 58}, 043506 (1998)
  [arXiv:astro-ph/9805260].
  %%CITATION = ASTRO-PH 9805260;%%
%----------------------------------------------------------------
\bibitem{cosmological-parameters}%
%\cite{Spergel:2003cb}
%\bibitem{Spergel:2003cb}
  D.~N.~Spergel {\it et al.}  [WMAP Collaboration],
%   ``First Year Wilkinson Microwave Anisotropy Probe (WMAP) Observations: Determination of Cosmological Parameters,''
  Astrophys.\ J.\ Suppl.\  {\bf 148}, 175 (2003)
  [arXiv:astro-ph/0302209].
  %%CITATION = ASTRO-PH 0302209;%%
%----------------------------------------------------------------
%\cite{Riess:2004nr}
%\bibitem{Riess:2004nr}
\bibitem{riess}
  A.~G.~Riess {\it et al.}  [Supernova Search Team Collaboration],
  %``Type Ia Supernova Discoveries at z>1 From the Hubble Space Telescope:
  %Evidence for Past Deceleration and Constraints on Dark Energy Evolution,''
  Astrophys.\ J.\  {\bf 607}, 665 (2004)
  [arXiv:astro-ph/0402512].
  %%CITATION = ASTRO-PH 0402512;%%
%----------------------------------------------------------------
%\cite{Bertolami:2004ic}
%\bibitem{Bertolami:2004ic}
\bibitem{bertolami}
  O.~Bertolami, A.~A.~Sen, S.~Sen and P.~T.~Silva,
  %``Latest Supernova data in the framework of Generalized Chaplygin Gas
  %model,''
  Mon.\ Not.\ Roy.\ Astron.\ Soc.\  {\bf 353}, 329 (2004)
  [arXiv:astro-ph/0402387].
  %%CITATION = ASTRO-PH 0402387;%%
%----------------------------------------------------------------
%\cite{Liddle:2006kn}
%\bibitem{Liddle:2006kn}
\bibitem{liddle}
  A.~R.~Liddle, P.~Mukherjee, D.~Parkinson and Y.~Wang,
  ``Present and future evidence for evolving dark energy,''
  arXiv:astro-ph/0610126.
  %%CITATION = ASTRO-PH 0610126;%%
%----------------------------------------------------------------




%------------------------------------------------------------------------------
\end{thebibliography}
\end{document}